\documentstyle[twoside,epsfig]{article}

\catcode`\@=11
\long\def\@makefntext#1{
\protect\noindent \hbox to 3.2pt {\hskip-.9pt  
$^{{\eightrm\@thefnmark}}$\hfil}#1\hfill}		%CAN BE USED 

\def\thefootnote{\fnsymbol{footnote}}
\def\@makefnmark{\hbox to 0pt{$^{\@thefnmark}$\hss}}	%ORIGINAL 
	
\def\ps@myheadings{\let\@mkboth\@gobbletwo
\def\@oddhead{\hbox{}
\rightmark\hfil\eightrm\thepage}   
\def\@oddfoot{}\def\@evenhead{\eightrm\thepage\hfil
\leftmark\hbox{}}\def\@evenfoot{}
\def\sectionmark##1{}\def\subsectionmark##1{}}

\oddsidemargin=\evensidemargin
\renewcommand{\thefootnote}{\fnsymbol{footnote}}

\newcounter{sectionc}\newcounter{subsectionc}\newcounter{subsubsectionc}
\renewcommand{\section}[1] {\vspace{12pt}\addtocounter{sectionc}{1} 
\setcounter{subsectionc}{0}\setcounter{subsubsectionc}{0}\noindent 
	{\tenbf\thesectionc. #1}\par\vspace{5pt}}
\renewcommand{\subsection}[1] {\vspace{12pt}\addtocounter{subsectionc}{1} 
	\setcounter{subsubsectionc}{0}\noindent 
	{\bf\thesectionc.\thesubsectionc. {\kern1pt \bfit #1}}\par\vspace{5pt}}
\renewcommand{\subsubsection}[1] {\vspace{12pt}\addtocounter{subsubsectionc}{1}
	\noindent{\tenrm\thesectionc.\thesubsectionc.\thesubsubsectionc.
	{\kern1pt \tenit #1}}\par\vspace{5pt}}
\newcommand{\nonumsection}[1] {\vspace{12pt}\noindent{\tenbf #1}
	\par\vspace{5pt}}

\newcounter{appendixc}
\newcounter{subappendixc}[appendixc]
\newcounter{subsubappendixc}[subappendixc]
\renewcommand{\thesubappendixc}{\Alph{appendixc}.\arabic{subappendixc}}
\renewcommand{\thesubsubappendixc}
	{\Alph{appendixc}.\arabic{subappendixc}.\arabic{subsubappendixc}}

\renewcommand{\appendix}[1] {\vspace{12pt}
        \refstepcounter{appendixc}
        \setcounter{figure}{0}
        \setcounter{table}{0}
        \setcounter{lemma}{0}
        \setcounter{theorem}{0}
        \setcounter{corollary}{0}
        \setcounter{definition}{0}
        \setcounter{equation}{0}
        \renewcommand{\thefigure}{\Alph{appendixc}.\arabic{figure}}
        \renewcommand{\thetable}{\Alph{appendixc}.\arabic{table}}
        \renewcommand{\theappendixc}{\Alph{appendixc}}
        \renewcommand{\thelemma}{\Alph{appendixc}.\arabic{lemma}}
        \renewcommand{\thetheorem}{\Alph{appendixc}.\arabic{theorem}}
        \renewcommand{\thedefinition}{\Alph{appendixc}.\arabic{definition}}
        \renewcommand{\thecorollary}{\Alph{appendixc}.\arabic{corollary}}
        \renewcommand{\theequation}{\Alph{appendixc}.\arabic{equation}}
        \noindent{\tenbf Appendix#1}\par\vspace{5pt}}
\newcommand{\subappendix}[1] {\vspace{12pt}
        \refstepcounter{subappendixc}
        \noindent{\bf Appendix \thesubappendixc. {\kern1pt \bfit #1}}
	\par\vspace{5pt}}
\newcommand{\subsubappendix}[1] {\vspace{12pt}
        \refstepcounter{subsubappendixc}
        \noindent{\rm Appendix \thesubsubappendixc. {\kern1pt \tenit #1}}
	\par\vspace{5pt}}

\newcommand{\textlineskip}{\baselineskip=13pt}
\newcommand{\smalllineskip}{\baselineskip=10pt}

\def\eightcirc{
\begin{picture}(0,0)
\put(4.4,1.8){\circle{6.5}}
\end{picture}}
\def\eightcopyright{\eightcirc\kern2.7pt\hbox{\eightrm c}} 

\newcommand{\copyrightheading}[1]
	{\vspace*{-2.5cm}\smalllineskip{\flushleft
	 }}

\def\abstracts#1#2#3{{
	\centering{\begin{minipage}{4.5in}\baselineskip=10pt\footnotesize
	\parindent=0pt #1\par 
	\parindent=15pt #2\par
	\parindent=15pt #3
	\end{minipage}}\par}} 

\renewenvironment{thebibliography}[1]
	{\frenchspacing
	 \ninerm\baselineskip=11pt
	 \begin{list}{\arabic{enumi}.}
        {\usecounter{enumi}\setlength{\parsep}{0pt}     
	 \setlength{\leftmargin 12.7pt}{\rightmargin 0pt} %FOR 1--9 ITEMS
         \setlength{\itemsep}{0pt} \settowidth
	{\labelwidth}{#1.}\sloppy}}{\end{list}}

\newcounter{itemlistc}
\newcounter{romanlistc}
\newcounter{alphlistc}
\newcounter{arabiclistc}

\newcommand{\fcaption}[1]{
        \refstepcounter{figure}
        \setbox\@tempboxa = \hbox{\footnotesize Fig.~\thefigure. #1}
        \ifdim \wd\@tempboxa > 5in
           {\begin{center}
        \parbox{5in}{\footnotesize\smalllineskip Fig.~\thefigure. #1}
            \end{center}}
        \else
             {\begin{center}
             {\footnotesize Fig.~\thefigure. #1}
              \end{center}}
        \fi}

\newcommand{\tcaption}[1]{
        \refstepcounter{table}
        \setbox\@tempboxa = \hbox{\footnotesize Table~\thetable. #1}
        \ifdim \wd\@tempboxa > 5in
           {\begin{center}
        \parbox{5in}{\footnotesize\smalllineskip Table~\thetable. #1}
            \end{center}}
        \else
             {\begin{center}
             {\footnotesize Table~\thetable. #1}
              \end{center}}
        \fi}

\def\@citex[#1]#2{\if@filesw\immediate\write\@auxout
	{\string\citation{#2}}\fi
\def\@citea{}\@cite{\@for\@citeb:=#2\do
	{\@citea\def\@citea{,}\@ifundefined
	{b@\@citeb}{{\bf ?}\@warning
	{Citation `\@citeb' on page \thepage \space undefined}}
	{\csname b@\@citeb\endcsname}}}{#1}}

\newif\if@cghi
\def\cite{\@cghitrue\@ifnextchar [{\@tempswatrue
	\@citex}{\@tempswafalse\@citex[]}}
\def\citelow{\@cghifalse\@ifnextchar [{\@tempswatrue
	\@citex}{\@tempswafalse\@citex[]}}
\def\@cite#1#2{{$\null^{#1}$\if@tempswa\typeout
	{IJCGA warning: optional citation argument 
	ignored: `#2'} \fi}}

\def\pmb#1{\setbox0=\hbox{#1}
	\kern-.025em\copy0\kern-\wd0
	\kern.05em\copy0\kern-\wd0
	\kern-.025em\raise.0433em\box0}

\def\fnt#1#2{\footnotetext{\kern-.3em
	{$^{\mbox{\scriptsize #1}}$}{#2}}}

\def\fpage#1{\begingroup
\voffset=.3in
\thispagestyle{empty}\begin{table}[b]\centerline{\footnotesize #1}
	\end{table}\endgroup}

\def\runninghead#1#2{\pagestyle{myheadings}
\markboth{{\protect\footnotesize\it{\quad #1}}\hfill}
{\hfill{\protect\footnotesize\it{#2\quad}}}}
\font\tenrm=cmr10
\font\tenit=cmti10 
\font\tenbf=cmbx10
\font\bfit=cmbxti10 at 10pt
\font\ninerm=cmr9

\font\eightrm=cmr8

\def\qed{\hbox{${\vcenter{\vbox{			%HOLLOW SQUARE
   \hrule height 0.4pt\hbox{\vrule width 0.4pt height 6pt
   \kern5pt\vrule width 0.4pt}\hrule height 0.4pt}}}$}}

\renewcommand{\thefootnote}{\fnsymbol{footnote}}	%USE SYMBOLIC FOOTNOTE

\def\theequation{\thesectionc.\arabic{equation}}	%FOR SETTING EQ.~(1.1)

\begin{document}
\runninghead{Multifractal and scaling properties in finance}
{Multifractal and scaling properties in finance}

\normalsize\textlineskip
\thispagestyle{empty}
\setcounter{page}{1}

\copyrightheading{}			%{Vol. 0, No.0 (1992) 000--000}
{{\em Int. J. Theor. Appl. Fin.}, Vol. 3, No.3 (2000) 361--364}

\vspace*{0.88truein}

\fpage{1}
\centerline{\bf MULTIFRACTAL FLUCTUATIONS}
\vspace*{0.035truein}
\centerline{\bf IN FINANCE}
\vspace*{0.37truein}
\centerline{\footnotesize FRANCOIS SCHMITT}
\vspace*{0.015truein}
\centerline{\footnotesize\it Dept. of Fluid Mechanics
VUB, 2 Pleinlaan}
\baselineskip=10pt
\centerline{\footnotesize\it B-1050 Brussels, Belgium}
\baselineskip=10pt
\centerline{\footnotesize\it francois@stro.vub.ac.be}
\vspace*{10pt}
\centerline{\footnotesize DANIEL SCHERTZER}
\vspace*{0.015truein}
\centerline{\footnotesize\it LMM, University of Paris VI}
\baselineskip=10pt
\centerline{\footnotesize\it 4, Place Jussieu, F-75005 Paris, France}
\vspace*{10pt}
\centerline{\footnotesize SHAUN LOVEJOY}
\vspace*{0.015truein}
\centerline{\footnotesize\it McGILL University, Physics Department}
\baselineskip=10pt
\centerline{\footnotesize\it 3600 University Street, Montreal H3A2T8, Canada}
\vspace*{1truein}

\abstracts{
We consider the structure functions $S^{(q)}(\tau)$, i.e.  the moments of
order $q$ of
the increments  $X(t+\tau)-X(t)$ of the Foreign Exchange rate
 $X(t)$
which give clear evidence of scaling ($S^{(q)}(\tau) \propto \tau^{\zeta(q)}$).
We demonstrate that the nonlinearity of the observed scaling exponent
$\zeta(q)$
is incompatible with monofractal additive
stochastic models usually introduced in finance:
Brownian motion, Levy processes and their truncated versions.
This nonlinearity correspond to multifractal intermittency yielded by
multiplicative processes. The non-analycity of $\zeta(q)$
corresponds to universal multifractals, which are
furthermore
able to produce ``hyperbolic'' pdf tails with an exponent $q_D >2$.
We argue that it is necessary to introduce stochastic evolution
equations which are compatible with this multifractal behaviour.
}{}{}

%\vspace*{10pt}
%\keywords{Multifractals, scaling, stable random variables}

\textlineskip			%) USE THIS MEASUREMENT WHEN THERE IS
\vspace*{12pt}			%) NO SECTION HEADING

%\vspace*{1pt}\textlineskip	%) USE THIS MEASUREMENT WHEN THERE IS
%\section{General Appearance}	%) A SECTION HEADING
%\vspace*{-0.5pt}

\textheight=7.8truein
\setcounter{footnote}{0}
\renewcommand{\thefootnote}{\alph{footnote}}

\section{The use of structure functions to discriminate models}
\noindent Financial markets display some common properties with fluid
turbulence, and their fluctuations are often characterized
as being ``turbulent''. Indeed, as for fluid turbulent
fluctuations, financial fluctuations display intermittency
 at all scales. In fluid turbulence, a cascade of energy flux
is known to occur from the large scale of injection to the small scales of
dissipation. Since the 1980's,
this cascade is mainly modeled by multiplicative cascades,$^1$ 
generically leading
to multifractal fields.$^{2,3}$

In finance, the picture of a cascade of information flowing from large-scale
investors to small scale ones has been proposed,$^{4-6}$
and several authors showed empirically that the fluctuations of various
financial time series possess multifractal statistics.$^{6-9}$ This
corresponds to
abandoning the classical Brownian motion picture, and even all other
models based on additive processes: fractionnal Brownian motion,
L\'evy and truncated L\'evy processes. Here we show how
structure function analysis is a simple yet powerful tool in
comparing the different models.

Assuming statistical time translational invariance, the
structure functios  $S^{(q)}(\tau)$,
i.e. the statistical moments of the increment 
of the Foreign Exchange rate $X(t)$
will depend only on the time lag  $\tau$, and according to
a power law if the process is scaling:
\begin{equation}
\label{eq2}
S^{(q)}(\tau)=< \vert X(t+\tau)-X(t) \vert ^q > \sim
S^{(q)}(T)\left(\frac{\tau}{T}\right)^{\zeta (q) }
\end{equation}
\noindent where $T$ is the fixed largest time scale of the system,
$<.>$ denotes statistical average (for non-overlapping increments
of length $\tau$), $q$ is the order of the moment
(we take here $q>0$), and $\zeta(q)$ is the scale invariant structure
function exponent. Structure function analysis corresponds in fact to
studying
``generalized'' average volatilities at scale $\tau$, since
only moments of order $1$ or
$2$ are usually used to define the volatility.  Furthermore, the
present analysis consists in analysing this generalized volatility
for all time scales.

The average of the fluctuations correspond to $q=1$, and $H=\zeta(1)$
is the so-called ``Hurst'' exponent characterizing the scaling non-conservation
of the mean. The second moment is linked to the slope $\beta$ of the
Fourier power spectrum: $\beta = 1+ \zeta(2)$. The
main property of a multifractal
processes is that it is characterized by a nonlinear $\zeta(q)$ function.$^{10}$
This function is convex, being
proportionnal to the second Laplace characteristic function
of the generator of the cascade.$^{1,3}$
Multifractals are the generic result of multiplicative cascades.
A continuous-scale limit of such processes leads to 
the family of log-infinitely
divisible distributions, among which are
the universal multifractals,$^{1}$ which  have a normal or Levy
generator, and for which:
\begin{equation}
\zeta (q) =qH-\frac{C_1}{\alpha-1}(q^{\alpha}-q)
\end{equation}
\noindent where $C_1 \leq d$ is an intermittency parameter, $d$ is the 
dimension of the space (here thus $d=1$) and 
$0 < \alpha \leq 2$ is the basic parameter which characterizes
the process; $\alpha=2$ corresponds to the log-normal distribution
(a normal generator).

On the other hand, additive models correspond to a linear
or bilinear
$\zeta(q)$.
Indeed, for Brownian motion (Bm) $\zeta(q)=q/2$, and
for fractionnal Brownian motion (fBm) $\zeta(q)=qH$, for a fractionnal
integration of order $H+1/2$ of a Gaussian noise. Thus a purely linear
$\zeta(q)$ function indicates Bm or fBm. We showed numerically in Ref.6
that several ARCH and GARCH models quickly converge
to giving $\zeta(q)=q/2$. We obtained also a bilinear expression
for the quite popular L\'evy-stable and
truncated L\'evy-stable processes$^{11-13}$:
$\zeta(q) = qH$ for$q  < 1/H$ and $= 1$ for $q \geq 1/H$,
where $H=1/\alpha$, and $\alpha$ is the L\'evy index
($0\leq \alpha \leq 2$). For $q\geq 1/H$ the above expression is valid for one
realization; when the number of realization increases, because of
the divergence of moments of order $\alpha$ of L\'evy processes,
 $\zeta(q)$ diverges (but its estimate on finite samples is always
finite$^{1,3}$).

It is important to note that multiplicative cascades generically
produce also hyperbolic tails leading to divergence of moments
of order $q_D$ which could be larger than $2$,$^{1}$
whereas additive processes are bounded to $q_D<2$.
Equation (1.2) is a nonlinear
behaviour obtained for $q \leq q_D$, whereas for $q>q_D$,
$\zeta(q)$ is linear, with a slope depending on the number of realizations
studied.$^{2}$

\section{Multifractal data analysis: an example}
\noindent We show here as a case study the analysis of a daily
US Dollar/French Franc exchange rate from 1 January 1979 to 30 November 1993.
This corresponds to 3680 data points, and a scaling of nearly three orders
of magnitude (but the analysis of intraday data showed that the
scaling of the financial fluctuations can go from several minutes to
several years). 

\begin{figure*}[t]
  \centerline{\psfig{figure=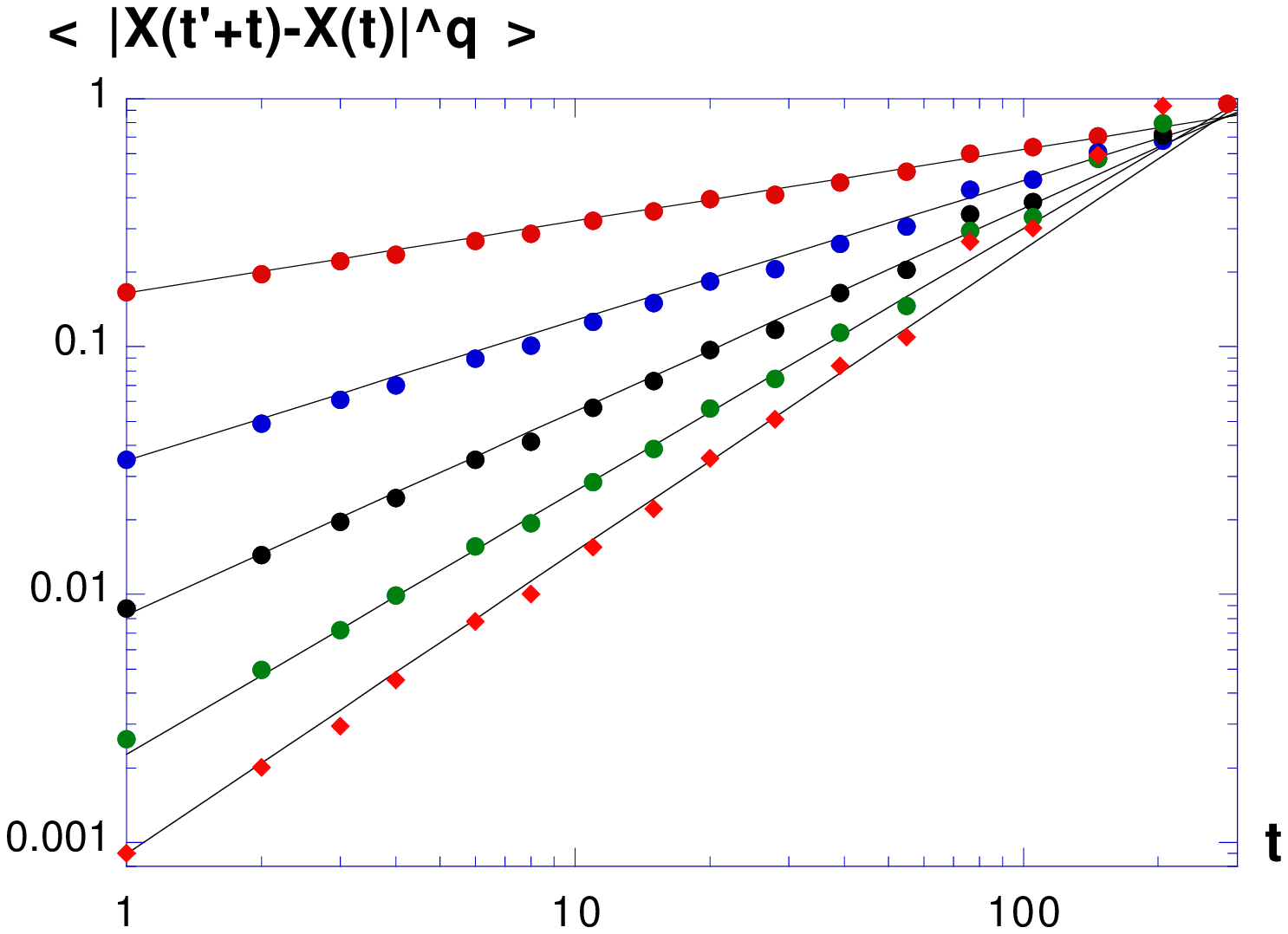,height=4.4cm}
              \psfig{figure=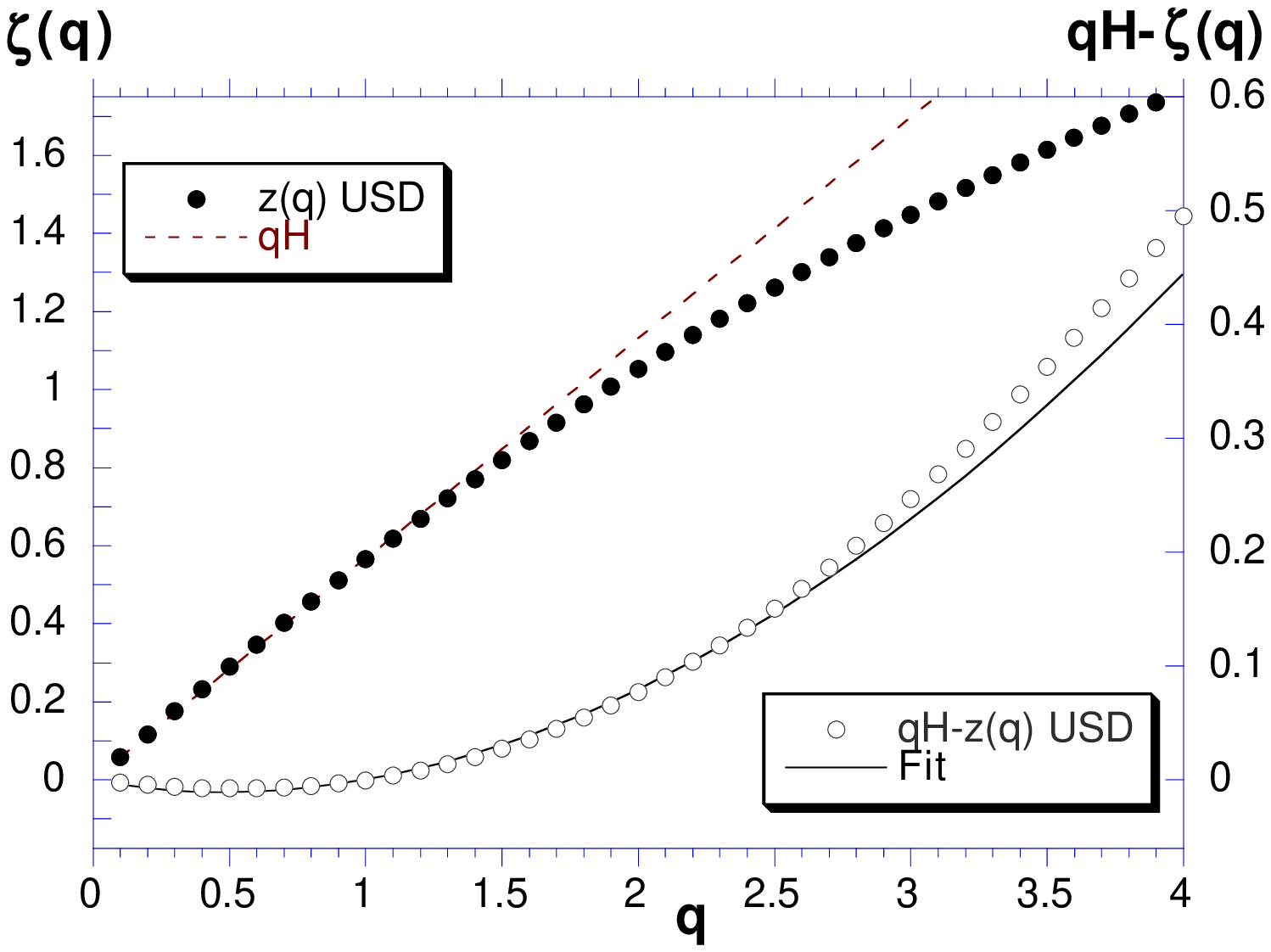,height=4.4cm}\\}
\vspace*{2pt}
  \caption{Left: Scaling of the structure functions in log-log plot
  for moments of order 0.5, 1., 1.5, 2. and 2.5 (from top to bottom);
Right: the resulting scaling exponent $\zeta(q)$ which is clearly nonlinear
(full dots), compared to the straight line $q\zeta(1)$ (dotted line). Also 
shown is the convex function $q\zeta(1)-\zeta(q)$ (open dots),
with a universal multifractal fit (continuous line).}
\end{figure*}

In Fig. 1a we show the structure functions
in log-log plot for different orders of moments. The straight lines
show that the scaling of Eq.(1) is very well
respected; we repeated this for moments up to $4.0$, with
a $0.1$ increment; only for moments larger than about $4.0$,
the scaling begins to be broken because
of the insufficient amount of data analyzed.
The resulting $\zeta(q)$ function is shown
in Fig. 1b: there is a clear nonlinearity; we also directly 
estimated
the scaling exponent of the nonlinear term 
$\tau ^{q H}/< (\Delta X_{\tau}) ^q >$, which is a convex function
 plotted on the same graph. We obtain 
the following values for several currencies
(slightly different
values for each currency were obtained, but nevertheless in each case 
the resulting
$\zeta(q)$ function was nonlinear): 
$H = .58 \pm 0.03$ and $\zeta(2)=1.06 \pm 0.05$,
and using specific analysis techniques, $C_1=.05 \pm .03$
and $\alpha =1.5 \pm .3$. We also obtained $q_D=3.0\pm.5$,
a value which is confirmed by an analysis on a much larger dataset.$^{14}$
This seems to indicate that, in general, FX data are
characterized by multifractal processes with a hyperbolic slope
of 3; a structure function analysis does indeed display
a nonlinear curve up to the third order moment, and then 
a straight line with a slope linear in
$-\ln N$, where $N$ is the number of datapoints used for the
statistical estimates.$^{6}$

The main application of this
new approach is predictability: past and present values of the time
series can be exploited in order to provide an optimal forecast.
This contradicts the Efficient Market Hypothesis,
which relies on memoryless models. 
As with all symmetry principles, in the absence of specific, strong scale
breaking mechanisms, we must assume that the scaling is unbroken and that
the small empirical deviations are due to poor statistics. The observed
nonlinearity of $\zeta(q)$ thus demonstrates that it is
multifractal.  Nevertheless, proposals have been made to either
intentionally$^{15}$ or inadvertently$^{16}$ drop the
scaling assumption and consider complex transient ("cross-over") regimes of
models which are only monoscaling in the limit.
However, let us emphasis that this corresponds to come back to a
non-scaling framework, and by consequence to face many theoretical
and practical difficulties.

\nonumsection{References}

\end{document}